\documentclass[useAMS]{mn2e}
\usepackage{amsmath}
\usepackage{textcomp}
\usepackage{amssymb}
\usepackage[pdftex]{graphicx}
\usepackage{amssymb}

\usepackage[margin=.8in, paperwidth=8.5in, paperheight=11in]{geometry}
\usepackage{multirow}
\usepackage{multicol}
\usepackage{epsfig}
\usepackage{graphicx}
\usepackage{esint}
\begin{document}

\title{Effects of Radiation Field Geometry on Line Driven Disc Winds}
\author[S. Dyda, D. Proga]
{\parbox{\textwidth}{Sergei~Dyda\thanks{sdyda@physics.unlv.edu}, Daniel Proga}\\
Department of Physics \& Astronomy, University of Nevada Las Vegas, Las Vegas, NV 89154
}

\date{\today}
\pagerange{\pageref{firstpage}--\pageref{lastpage}}
\pubyear{2018}

\label{firstpage}

\maketitle

\begin{abstract}
We study line driven winds for models with different radial intensity profiles: standard Shakura-Sunyaev radiating thin discs, uniform intensity discs and truncated discs where driving radiation is cutoff at some radius. We find that global outflow properties depend primarily on the total system luminosity but truncated discs can launch outflows with $\sim 2$ times higher mass flux and $\sim 50\%$ faster outflow velocity than non-truncated discs with the same total radiation flux. Streamlines interior to the truncation radius are largely unaffected and carry the same momentum flux as non-truncated models whereas those far outside the truncation radius effectively carry no outflow because the local radiation force is too weak to lift matter vertically away from the disc. Near the truncation radius the flow becomes more radial, due to the loss of pressure/radiation support from gas/radiation at larger radii. These models suggest that line driven outflows are sensitive to the geometry of the radiation field driving them, motivating the need for self-consistent disc/wind models.

\end{abstract}

\begin{keywords} 
radiation: dynamics - hydrodynamics - stars:massive - stars: winds, outflows - quasars: general - X-rays: galaxies 
\end{keywords}

\section{Introduction}
Outflows are ubiquitous for many compact object systems with accretion discs, including cataclysmic variables (CVs), X-ray binaries (XRBs) and actve galactic nuclei (AGN). A possible mechanism for launching these outflows is radiation pressure due to spectral lines, so called line driving (Lucy and Solomon 1970, Castor, Abbott and Klein 1975). Simulating line driven winds is challenging because it requires both correctly modeling the microphysics governing the ionization state of the gas (Owocki, Cranmer \& Gayley 1996), and the macrophysics governing the hydrodynamics. Both are affected by the radiation field, the former most sensitive to the ionizing flux, and the latter to the flux in UV where there are very many spectral lines. Correctly treating the radiation field is thus critical in modeling line driven winds.

For systems with accretion discs an important source of UV radiation is the disc itself. The earliest stationary, thin disc, models assumed that accretion energy, dissipated by an effective viscosity, is radiated as blackbody radiation (Shakura \& Sunyaev 1973). This standard accretion disc model allows one to calculate the temperature and spectral energy distribution (SED) for the entire disc. The disc structure and therefore the radiation field as well, are modified when one accounts for outflows acting as a sink of mass and angular momentum in the disc. This has been shown in both the context of CVs (Knigge 1999) and AGN (Laor \& Davis 2014) where radiation is thought to drive outflows. Further, in the context of AGN, microlensing observations suggest that optical emission regions of the lensed quasars are typically larger than expected from basic thin-disk models by factors of $\sim 3-30$ (e.g. Pooley et al. 2007). Both theory and observations suggest that the standard disc model is an incomplete description of the radiation field near accretion discs.

Line driven disc wind simulations have shown that to first approximation global outflow properties (mass flux, outflow velocity and wind opening angle) depend only on the total system luminosity (Proga, Stone \& Drew 1998, hereafter PSD98). However the structure of the flow is sensitive to the geometry of the radiation field. Systems where radiation is sourced primarily from the disc have been found to have small scale structure in both 2D axisymmetric (Proga, Stone \& Drew 1999) and 3D (Dyda \& Proga 2018a, b, hereafter DP18a and DP18b) simulations. This suggests that to correctly model line driven winds we must move beyond the radiating thin disc, and compute the radiation field self consistently.   

Previous disc wind simulations used a frequency integrated disc intensity. Most spectral lines are in the UV so these models implicitly assume sufficient UV flux to drive the wind. If discs locally emit like blackbodies, there will be a range of radii for which the local disc spectrum peaks in the UV. Interior to some radii, where $T \gtrsim 30 \ 000 \ \rm{K}$ and exterior to some radii where $T \lesssim 7 \ 000 \ \rm{K}$ the spectra will peak outside  the $10 \ \rm{nm} \lesssim \lambda_{\rm{UV}} \lesssim 400 \ \rm{nm}$ UV wavelengths, significantly limiting the flux of photons available to drive a wind with radiation pressure due to lines. Proga et al. (2004) investigated this issue in the case of AGN where they showed that even if the UV to X-ray flux ratio was lowered to 1:9, line driven winds could still be launched as when this ratio is 1:1 as in Proga et al. (2000). The full problem of multi-frequency radiation transfer is currently computationally intractable. Therefore as a first step we investigate the effects of changing the radial intensity profile as a proxy for a diminished UV photon flux. Our goal is to correlate the geometry of the radiation field and the geometry of the outflow, to investigte the disc/wind connection in line driven disc winds.

In our previous studies of line driven disc winds we assumed the disc radiation field was sourced by a standard accretion disc, locally emitting like a blackbody. We consider two modificaions to this scenario. In one set of models, the intensity profile is as in the standard accretion disc, up until some outer truncation radius, $r_{\rm{c}}$, where the disc is assumed too cold to supply sufficient UV photons and the radiation field is exponentially suppressed. In the other set of models, we assume a constant radial intensity profile interior to the cutoff and keep the total disc luminosity fixed as we vary the truncation radius. We investigate how global outflow properties as well as the geometry of the flow are altered by varying the truncation radius, while controlling for the fact that to first approximation the solution is determined by the total system luminosity.

The structure of this paper is as follows. In Section \ref{sec:numerical} we briefly describe our numerical methods and code. In Section \ref{sec:truncation} we describe the effects of varying the truncation radius on outflow properties. In Section \ref{sec:luminosity}, using uniform intensity disc models, we show that changes due to introducing a truncation radius cannot be explained solely due to a change in total disc luminosity. In Section \ref{sec:discussion} we discuss some observational implications for our models, describe some limitations of our approach and sketch some possible future directions for this work.

\section{Numerical Methods}
\label{sec:numerical}
We performed all numerical simulations with the publicly available MHD code \textsc{Athena++} (Gardiner \& Stone 2005, 2008). The basic physical setup is a gravitating, central object surrounded by a thin, luminous accretion disc. The accretion disc sources a radiation field to drive the gas that is optically thin to the continuum.  Our basic physical setup is the same as described in DP18a,b, but we outline our method here, as well as any differences from our previous work below. 

\subsection{Basic Equations}
\label{sec:setup}

The basic equations for single fluid hydrodynamics driven by a radiation field are
\begin{subequations}
\begin{equation}
\frac{\partial \rho}{\partial t} + \nabla \cdot \left( \rho \mathbf{v} \right) = 0,
\end{equation}
\begin{equation}
\frac{\partial (\rho \mathbf{v})}{\partial t} + \nabla \cdot \left(\rho \mathbf{vv} + \mathbf{P} \right) = - \rho \nabla \Phi + \rho \mathbf{F}^{\rm{rad}},
\label{eq:momentum}
\end{equation}
\begin{equation}
\frac{\partial E}{\partial t} + \nabla \cdot \left( (E + P)\mathbf{v} \right) = -\rho \mathbf{v} \cdot \nabla \Phi + \rho \mathbf{v} \cdot \mathbf{F}^{\rm{rad}} ,
\label{eq:energy}
\end{equation}
\label{eq:hydro}%
\end{subequations}
where $\rho$, $\mathbf{v}$ are the fluid density and velocity respectively and $\mathbf{P}$ is a diagonal tensor with components P the gas pressure. For the gravitational potential, we use $\Phi = -GM/r$ and $E = 1/2 \rho |\mathbf{v}|^2 + \mathcal{E}$ is the total energy where $\mathcal{E} =  P/(\gamma -1)$ is the internal energy. The total radiation force per unit mass, $\mathbf{F}^{\rm{rad}}$, primarily imparts momentum on the gas (equation \ref{eq:momentum}), but we include the work done by it on the gas (equation \ref{eq:energy}) for self consistency. The isothermal sound speed is $a^2 = P/\rho$ and the adiabatic sound speed $c_s^2 = \gamma a^2$. We use the equation of state $P = k \rho^{\gamma}$ with $\gamma = 1.01$.   The temperature is then $T = (\gamma -1)\mathcal{E}\mu m_{\rm{p}}/\rho k_{\rm{b}}$ where $\mu = 0.6$ is the mean molecular weight and other symbols have their standard meaning.

\subsection{Radiation Force}
\label{sec:radiation_force}

\begin{figure}
                \centering
                \includegraphics[width=0.45\textwidth]{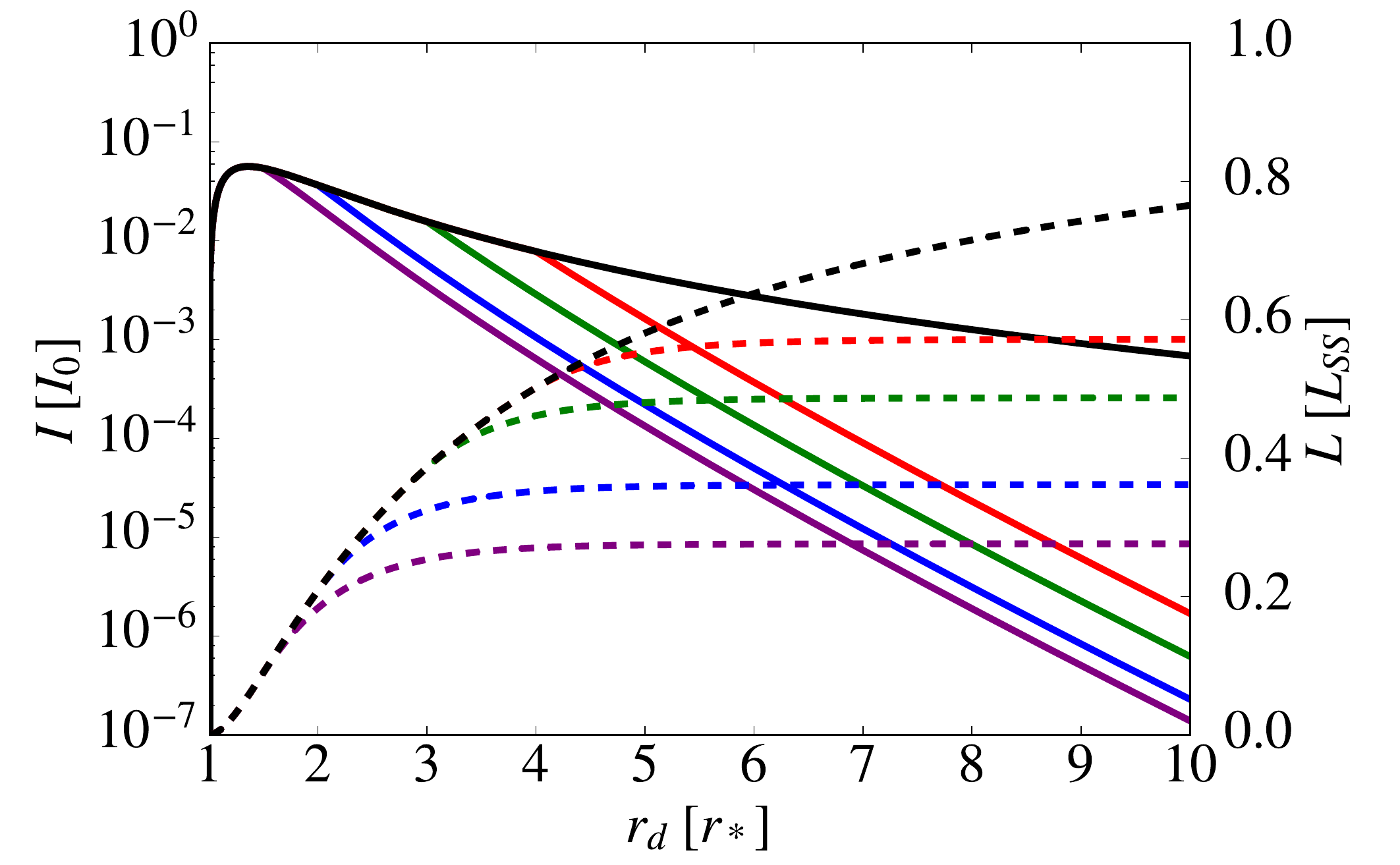}
        \caption{Intensity $I$ as a function of radial distance $r_d$ (solid lines) and luminosity interior to the indicated radius $L(r_d)$ normalized to Shakura-Sunyaev disc luminosity $L_{SS}$ (dashed lines) for standard discs truncated at radii $r_c = 1.5 r_*$ (purple), $2 r_*$ (blue), $3 r_*$ (green), $\ 4 r_*$ (red) and $\ 30 r_*$ (black).}
\label{fig:intensity}
\end{figure} 

We assume a time independent radiation field is sourced by an axisymmetric, geometrically thin accretion disc along the midplane. Every point in the wind experiences a radiation force
\begin{equation}
\mathbf{F}^{\rm{rad}} = \mathbf{F}^{\rm{rad}}_e + \mathbf{F}^{\rm{rad}}_{L},
\end{equation}  
which is a sum of the contributions due to electron scattering $\mathbf{F}^{\rm{rad}}_e$ and line driving $\mathbf{F}^{\rm{rad}}_{L}$. We explain in detail how to compute the radiation force in DP18a,b. It suffices to say that in assuming that the wind is optically thin to the continuum, the radiation force is proportional to the disc intensity.  The frequency integrated intensity of a Shakura-Sunyaev disc with inner radius $r_*$, as a function of radial distance $r_d$, is
\begin{align}
I_{\rm{SS}}(r_d) &=  I_0 \Gamma_d  \left( \frac{r_*}{r_d} \right)^3 \left[ 1 - \left(\frac{r_*}{r_d} \right)^{1/2}\right],
\label{eq:intensity_SS} 
\end{align} 
where the disc Eddington number
\begin{equation}
\Gamma_d = \frac{\dot{M}_{\rm{acc}} \sigma_e}{8 \pi c r_*},
\end{equation}
where $\dot{M}_{\rm{acc}}$ is the accretion rate in the disc (e.g., Pringle 1981), $c$ the speed of light, $\sigma_e$ the Thompson cross section and the fiducial intensity is
\begin{equation}
I_0 = \frac{3}{\pi} \frac{GM}{r_*^2}\frac{c}{\sigma_e}.
\end{equation} 
It is also useful from a theoretical context to consider a uniform intensity disc where
\begin{align}
I_{\rm{U}}(r_d) &= I_0 \Gamma_d.  
\label{eq:intensity_uniform} 
\end{align} 
We truncate the radiation field of the Shakura-Sunyaev or uniform intensity disc by introducing a cutoff radii $r_{c}$, where the disc is assumed to be too cold to radiate in the UV band where radiation can drive line driven winds,
\begin{equation}
I(r_d) = \begin{cases}
                                   I & r_d \leq r_c \\
  				I \exp \left\{ \frac{(r_c - r_d)}{r_*}\right\}  & r_d > r_c
  		\end{cases}
\label{eq:intensity}
\end{equation}
where the intensity $I$ is given by eq. (\ref{eq:intensity_SS}) or eq. (\ref{eq:intensity_uniform}) respectively. In Fig \ref{fig:intensity} we plot the intensity profile $I$ as a function of disc radius $r_d$ for each of these models. We also plot the luminosity interior to a given radius
\begin{equation}
L(r) = 2 \pi \int_{r_*}^{r} I(r') \ r' dr', 
\end{equation}
normalized to the Shakura-Sunyaev disc luminosity, $L_{SS} = L(\infty)$ where the intensity $I = I_{SS}$.

\section{Results}
\label{sec:results}

\begin{table*}
\begin{center}
    \begin{tabular}{| l | c | c | c| c |c| c | c | c| c|}
    \hline \hline
        &\multicolumn{3}{c}{Parameters} & \multicolumn{3}{c}{Global Properties} & \multicolumn{3}{c}{Streamline Properties} \\
Model	&	$\Gamma_d$	& $r_c \ [r_{*}]$	& $L \ M_{\rm{max}} \ [L_{Edd.}]$ 	&$\dot{M} \ [\rm{M_{\odot}/y}] $	&$v_{\rm{max}} \ [\rm{km/s}]$	&$\omega \ [^{\circ}]$ &$\bar{\rho} \ [\rm{g \ cm^{-3}}] $	&$\bar{v}_r \ [\rm{km/s}]$	&$\Delta \omega \ [^{\circ}]$  \\ \hline \hline
A	&$3.76 \times 10^{-4}$	&  $30$   &  1.5 & $2.3 \times 10^{-13}$ &  1,700 & 42 & $1.9 \times 10^{-18}$ & 1,600 &  8 \\
B	&$1.18 \times 10^{-3}$	&  $30$   &  4.7 & $7.8 \times 10^{-12}$ &  4,200 & 55 & $5.6 \times 10^{-15}$ & 2,000 & 14 \\
P	&$3.76 \times 10^{-3}$	&  $30$   & 15.1 & $9.1 \times 10^{-11}$ &  8,400 & 68 & $6.6 \times 10^{-14}$ & 1,700 & 17 \\
R	&$1.18 \times 10^{-2}$	&  $30$   & 47.4 & $5.4 \times 10^{-10}$ & 14,800 & 75 & $3.2 \times 10^{-13}$ & 2,200 & 22 \\ \hline
U4	&$9.44 \times 10^{-5}$	&  $4$ 	  & 15.6 & $1.9 \times 10^{-10}$ &  8,000 & 68 & $1.9 \times 10^{-13}$ & 1,200 & 10 \\ 
U3	&$1.48 \times 10^{-4}$	&  $3$ 	  & 15.6 & $1.7 \times 10^{-10}$ &  9,100 & 71 & $1.6 \times 10^{-13}$ & 1,800 & 18 \\
U2	&$2.62 \times 10^{-4}$	&  $2$ 	  & 15.6 & $1.6 \times 10^{-10}$ & 10,100 & 71 & $9.9 \times 10^{-14}$ & 2,300 & 18 \\
U1.5	&$3.78 \times 10^{-4}$	&  $1.5$  & 15.6 & $1.5 \times 10^{-10}$ & 10,400 & 71 & $7.7 \times 10^{-14}$ & 2,500 & 17 \\\hline
B4	&$1.18 \times 10^{-3}$	&  $4$	  &  3.0 & $4.3 \times 10^{-12}$ &  4,300 & 55 & $1.8 \times 10^{-15}$ & 2,400 & 17 \\
B3	&$1.18 \times 10^{-3}$	&  $3$	  &  2.5 & $2.9 \times 10^{-12}$ &  4,100 & 52 & $8.1 \times 10^{-16}$ & 2,600 & 26 \\
B2	&$1.18 \times 10^{-3}$	&  $2$	  &  1.9 & $1.2 \times 10^{-12}$ &  3,600 & 47 & 	-     	      &   -   & -  \\ 
B1.5	&$1.18 \times 10^{-3}$	&  $1.5$  &  1.4 & $4.7 \times 10^{-13}$ &  3,000 & 42 &         -             &   -   & -  \\ \hline
P4	&$3.76 \times 10^{-3}$	&  $4$	  &  9.4 & $5.8 \times 10^{-11}$ &  8,100 & 64 & $3.1 \times 10^{-14}$ & 2,500 & 26 \\
P3	&$3.76 \times 10^{-3}$	&  $3$	  &  8.0 & $4.3 \times 10^{-11}$ &  7,600 & 64 & $2.9 \times 10^{-14}$ & 2,200 & 32 \\
P2	&$3.76 \times 10^{-3}$	&  $2$	  &  6.0 & $2.4 \times 10^{-11}$ &  7,200 & 64 & $9.1 \times 10^{-15}$ & 3,000 & 32 \\ 
P1.5	&$3.76 \times 10^{-3}$	&  $1.5$  &  4.6 & $1.3 \times 10^{-11}$ &  6,400 & 61 & $3.9 \times 10^{-15}$ & 3,100 & 29 \\ \hline
R4	&$1.18 \times 10^{-2}$	&  $4$	  & 29.6 & $5.0 \times 10^{-10}$ & 14,500 & 75 & $4.8 \times 10^{-13}$ & 1,600 & 32 \\
R3	&$1.18 \times 10^{-2}$	&  $3$	  & 25.2 & $3.9 \times 10^{-10}$ & 13,600 & 75 & $2.3 \times 10^{-13}$ & 2,700 & 28 \\
R2	&$1.18 \times 10^{-2}$	&  $2$	  & 18.7 & $2.5 \times 10^{-10}$ & 12,200 & 64 & $9.1 \times 10^{-15}$ & 3,000 & 32 \\ 
R1.5	&$1.18 \times 10^{-2}$	&  $1.5$  & 14.3 & $1.4 \times 10^{-10}$ & 10,800 & 71 & $3.7 \times 10^{-14}$ & 3,900 & 34 \\\hline \hline

    \end{tabular}
\end{center}
\caption{Summary of wind properties for different disc wind models, including global properties (mass flux $\dot{M}$, maximum velocity $v_{\rm{max}}$ and opening angle $\omega$) and fast-stream properties (average density $\bar{\rho}$, density weighted average velocity $\bar{v}_r$ and angular size $\Delta \omega$ at the outer boundary for parts of the wind between streamlines with footpoints at $r = 1.5 \ r_*$ and $r = 3.0 \ r_*$). Models A, B, P and R are fiducial Shakura-Sunyaev models (see eq. \ref{eq:intensity_SS}) with no cutoff. Models A and B are described more fully in DP18b. Model U is a uniform intensity disc where total luminosity is kept fixed as the truncation radius is varied (see eq. \ref{eq:intensity_uniform}). The truncation radius, in units of $r_*$, follows the name of the corresponding fiducial model.}
\label{tab:summary}
\end{table*}

To study the effects of the radiation field on the flow geometry, we perform a series of disc wind simulations. The fiducial models (A, B, P \& R) correspond to standard radiating discs (see eq. \ref{eq:intensity_SS}) with Eddington fraction $3.76 \ \times 10^{-4} \leq \Gamma_d \leq 1.18 \ \times 10^{-2}$, where the disc is cutoff at $r = 30 \ r_*$ for computational reasons. For the three highest Eddington fraction models ($\Gamma_d = 1.18 \times 10^{-3}$, $3.76 \times 10^{-3}$ and $1.18 \times 10^{-2}$), we simulate discs where the cutoff radius $1.5 \ r_* \leq r_c \leq 4 \ r_*$ (see eq. \ref{eq:intensity}), which we denote by the letter corresponding to the model fiducial luminosity followed by the cutoff radius, i.e B1.5, ..., B4. Finally we perform simulations with a uniform intensity profile (see eq. \ref{eq:intensity_uniform}), where the cutoff again varies from $1.5 \ r_* \leq r_c \leq 4 \ r_*$ which we denote by U1.5, ..., U4.

In Table \ref{tab:summary} we list all our model parameters, including the Eddington fraction $\Gamma_d$, truncation radius $r_{\rm{c}}$ and total disc luminosity multiplied by the maximum force multiplier (maximum effective number of lines) $L M_{\rm{max}}$. For each model we summarize the global outflow properties (mass flux $\dot{M}$, maximum outflow velocity $v_{\rm{max}}$ and opening angle $\omega$) and for the flow between streamlines with footpoints fixed at $r = 1.5 \ r_*$ and $r = 3.0 \ r_*$ we show the mean density $\bar{\rho}$, density weighted mean velocity $\bar{v}_r$ and angle between the streamlines $\Delta \omega$ at the outer boundary.       

\subsection{Truncation Effects}
\label{sec:truncation}

\begin{figure*}
                \centering
                \includegraphics[width=\textwidth]{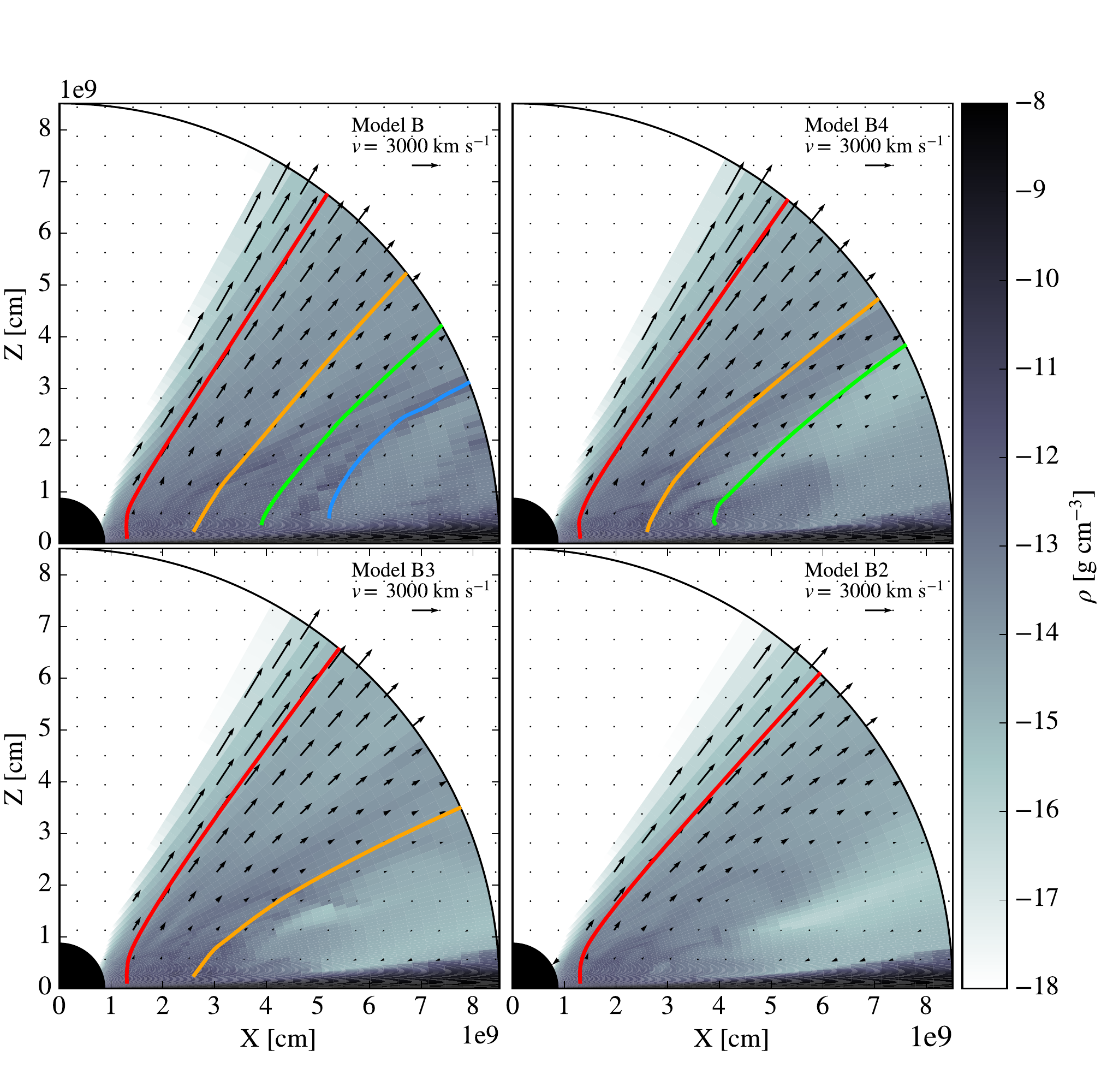}
        \caption{Time averaged density profile $\bar{\rho}$ (gray contours) and poloidal velocity field $\bar{\mathbf{v}}_p$ (black vectors) for $900 \ \rm{s} \leq t \leq 1000 \ \rm{s}$ for the different radiation models. The colored lines indicate streamlines with footpoint at $r = 1.5 \ r_*$ (red), $3.0 \ r_*$ (orange), $4.5 \ r_*$ (green) and $6.0 \ r_*$ (blue). We only plot streamlines that exit the computational domain. Streamlines interior to the truncation radius are largely unaffected whereas those exterior no longer carry mass out the simulation domain. Streamlines near the truncation radius become more radial, launching a less dense but faster flow.}
\label{fig:time_averaged_winds_B}
\end{figure*} 

\begin{figure*}
                \centering
                \includegraphics[width=\textwidth]{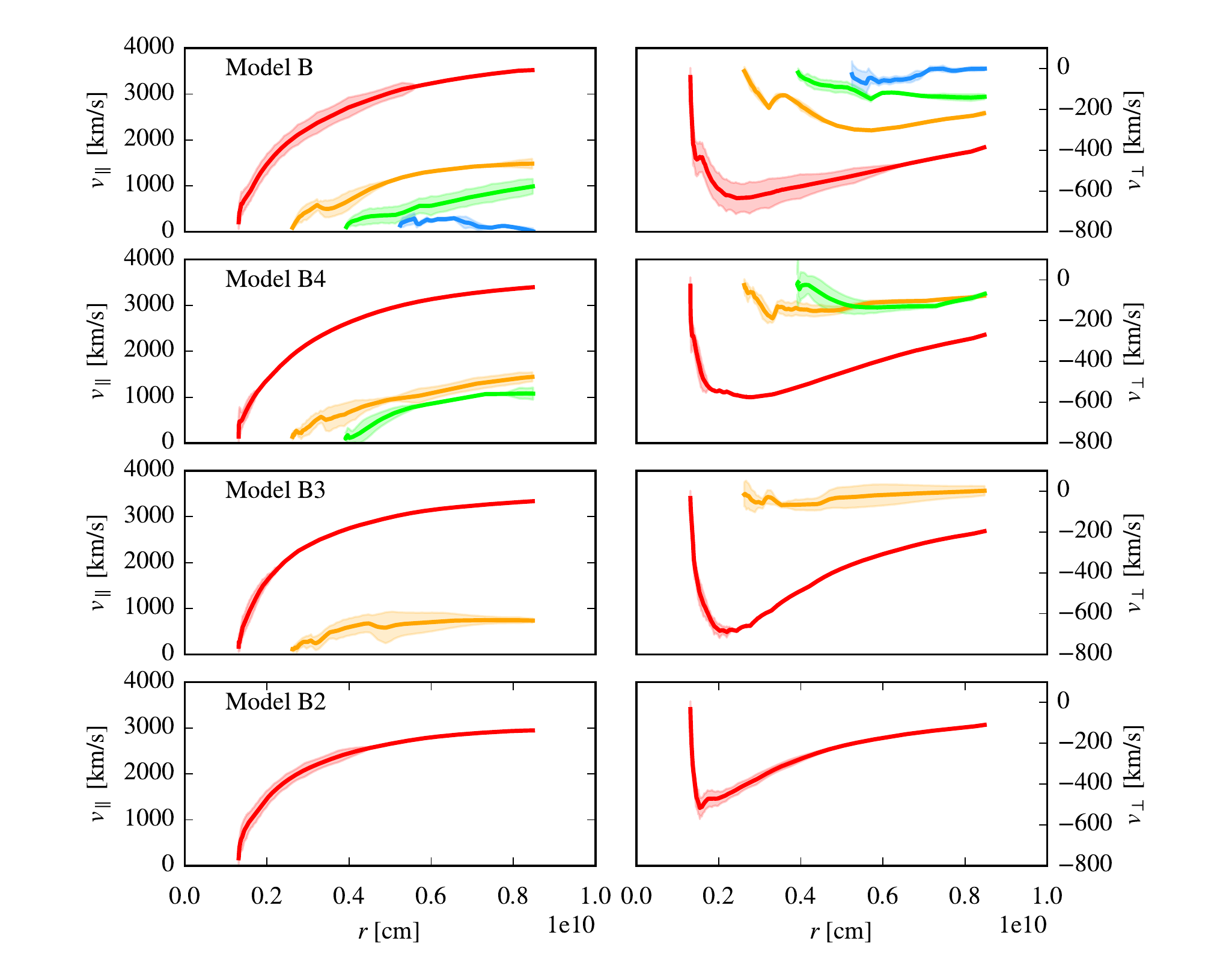}
        \caption{Velocity parallel $v_{\parallel}$ (left panels) and perpendicular $v_{\perp}$ (right panels) to time-averaged streamlines shown in Fig. \ref{fig:time_averaged_winds_B} (colored lines) and the corresponding velocity dispersion (shaded region). Outflow velocity decreases very slightly with decreasing truncation radius until the footpoint is interior to the truncation radius and velocity is strongly suppressed. Note the different scales for $v_{\parallel}$ and $v_{\perp}$.}
\label{fig:v_stream}
\end{figure*} 

\begin{figure}
                \centering
                \includegraphics[width=0.45\textwidth]{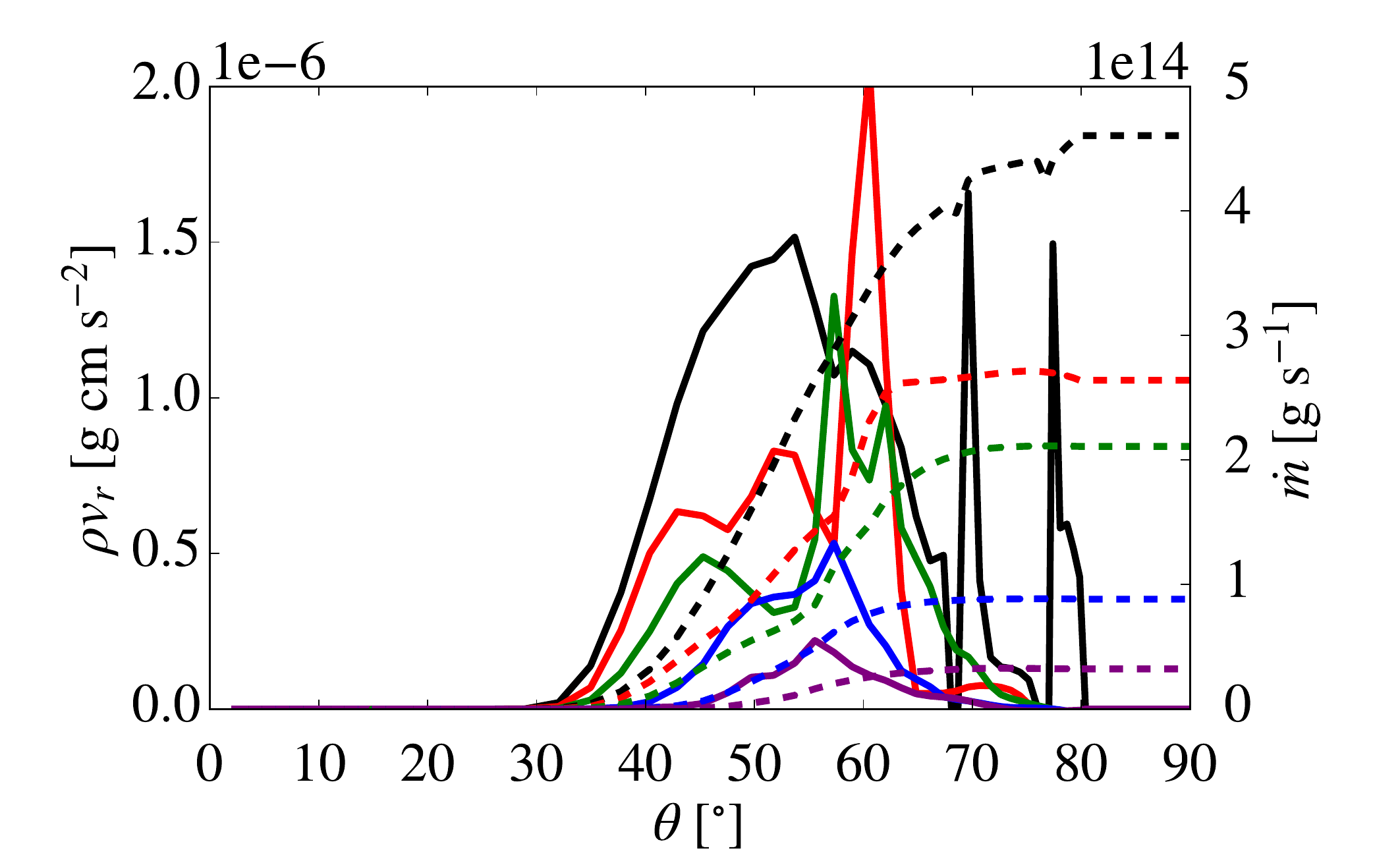}
                \includegraphics[width=0.45\textwidth]{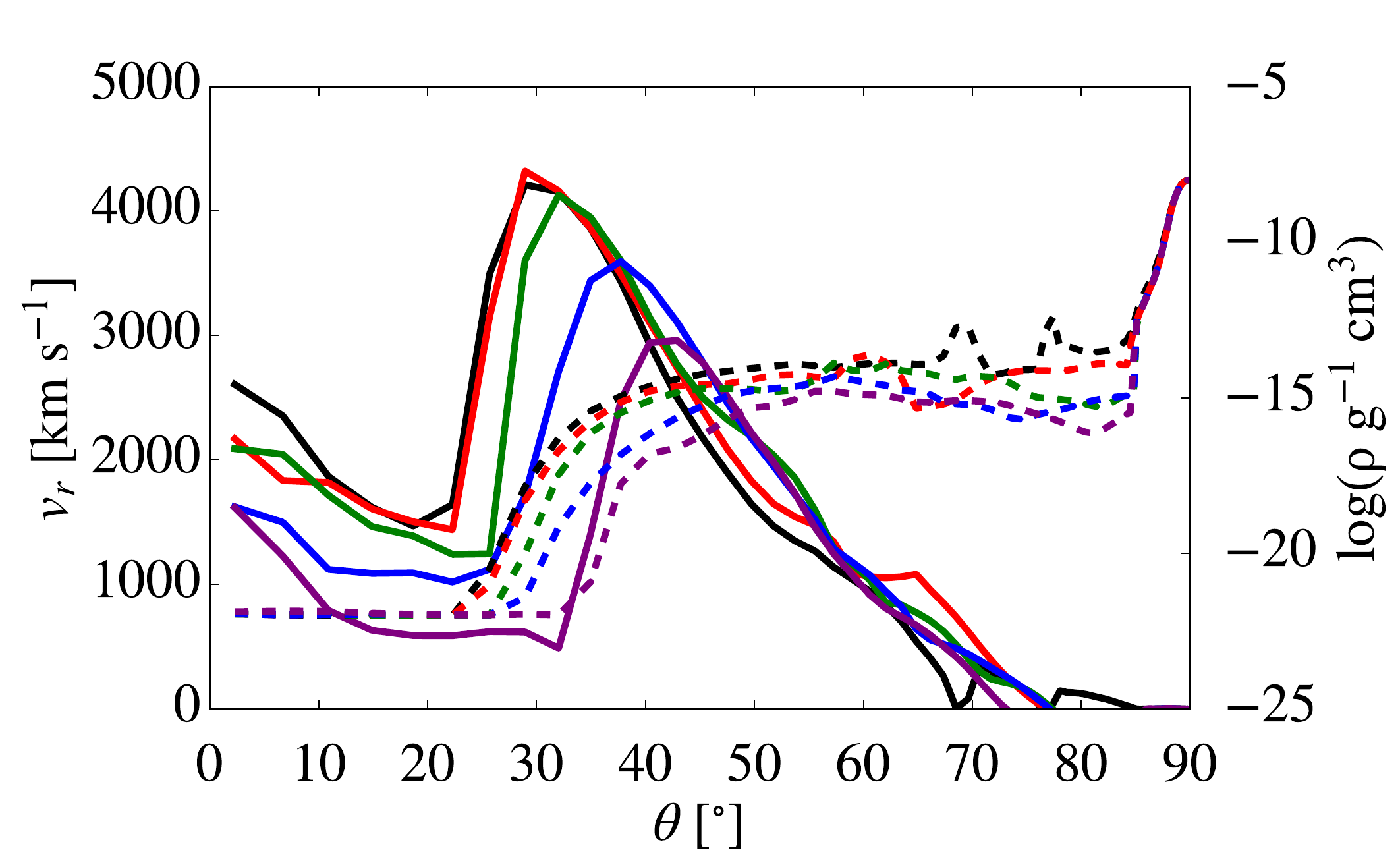}
        \caption{Time-averaged solutions over $900 \ \rm{s} \leq t \leq 1000 \ \rm{s}$ for models B (black), B4 (red), B3 (green), B2 (blue) and B1.5 (purple) at the outer boundary $r = r_o$. \textit{Top -} Momentum flux $\rho v_r$ (solid lines) and $\theta-$integrated mass flux $\dot{m}$ (dashed lines) as a function of $\theta$.  \textit{Bottom -} Velocity $v_r$ (solid lines) and density $\rho$ (dashed lines) as a function of $\theta$.}
\label{fig:outflow_avg}
\end{figure} 

\begin{figure}
                \centering
                \includegraphics[width=0.45\textwidth]{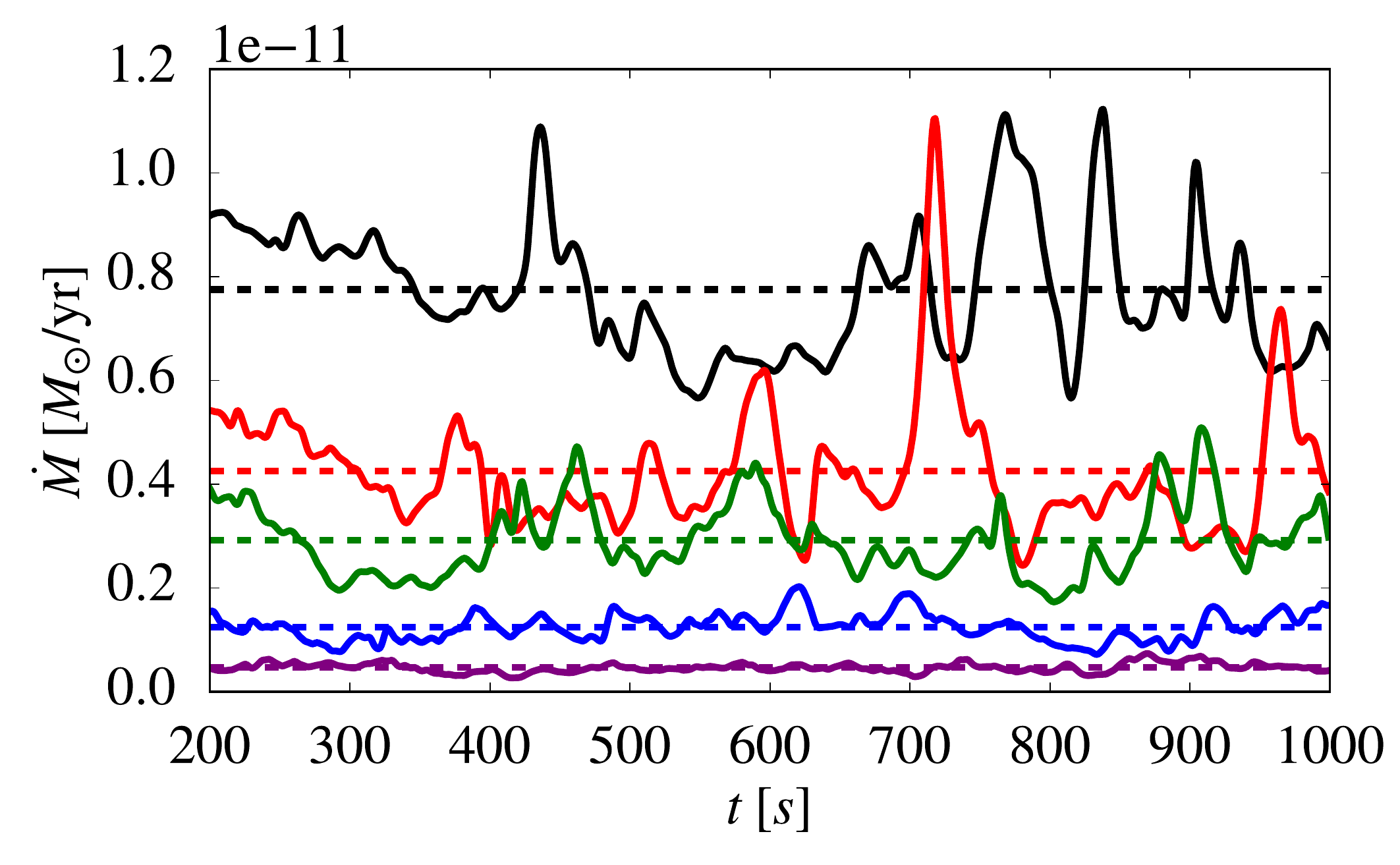}
        \caption{Mass flux $\dot{M}$ at the outer boundary as a function of time (solid lines) and the time average over $200 \ \rm{s} \leq t \leq 1000 \ \rm{s}$ (dashed line). Time variability is $\sim 20\%$, irrespective of the model.}
\label{fig:mdot_summary}
\end{figure} 

We study the effects of radiation field geometry on the flow by varying the radius at which disc intensity is truncated, $r_c$. We choose Model B from DP18b as our fiducial model, and vary the cutoff radius $r_{\rm{c}} = 1.5 r_*, 2 r_*$, $3 r_*$ and $4 r_* $.  

In Fig. \ref{fig:time_averaged_winds_B}, we plot the time averaged density profile $\bar{\rho}$ (gray contours) and poloidal velocity field $\bar{\mathbf{v}}_p$ (black vectors) for $900 \ \rm{s} \leq t \leq 1000 \ \rm{s}$ for fiducial Model B and the related truncated disc models. The colored lines indicate streamlines with footpoint at $r = 1.5r_*$ (red), $3.0 r_*$ (orange), $4.5 r_*$ (green) and $6.0r_*$ (blue). The basic structure of the winds are very similar, with a low density polar funnel at small $\theta$, a fast-stream carrying most of the mass flux and hydrostatic disc. The wind opening angle is only weakly dependent on total disc luminosity, decreasing from $\omega = 55^{\circ}$ for model B to $\omega = 42^{\circ}$ for model B1.5. This is in line with our expectations whereby the truncated disc models opening angles are bounded from above by non-truncated models (B) and have similar opening angles to non-truncated models with similar total luminosity (model A).

These disc wind solutions are non-stationary, so to investigate the effects of disc truncation on time-dependence, we consider the velocity dispersion. In Fig. \ref{fig:v_stream}, we plot the velocity parallel $v_{\parallel}$ and perpendicular $v_{\perp}$ to the time-averaged streamlines shown in Fig. \ref{fig:time_averaged_winds_B}, (colored lines) and the corresponding velocity dispersion (shaded region). We only plot the streamlines that exit the simulation domain, indicating that the flow is relatively stable. Truncating the disc has little effect on streamlines \emph{interior} to $r_c$. For example, all models have the streamline starting at $r = 1.5 r_*$ (red line) exiting the simulation domain. The outflow velocity is relatively constant with $v_{\parallel} \approx 3500 \rm{km/s}$ (model B) to $v_{\parallel} \approx 3000 \rm{km/s}$ (model B2). There is little change in the structure of the fast-stream, because its structure is primarily determined by streamlines in the innermost part of the disc. This explains why the mass flux, which is primarily determined by the fast-stream, is relatively unchanged in the truncated disc models. Streamlines \emph{exterior} to the cutoff radius become unstable and no longer carry mass out from the simulation domain so the line driving does not launch an outflow and only puffs up the disc. Near the disc, geometric foreshortening is important and the radiation force is approximately set by the local disc field. Therefore for $r > r_c$, the radiation force is too weak to vertically lift the gas above the disc where it experiences a force from the interior part of the disc. The exterior streamlines (not shown) typically travel radially inward, indicating that the radiation force is unable to balance gravity.

Streamlines originating at radii smaller but close to $r_c$ are the most affected by the truncated disc. Gas is lifted upwards, and because gas exterior to it is not being launched, it feels a pressure gradient (relative to the non-truncated disc) that makes the streamline more radial. The flow is thus able to diverge more as the gas expands out into the region below the fast-stream. There is also a reduced radiation force from exterior parts of  the disc which would also tend to make the flow more vertical. We see this most clearly from the streamline with footpoint at $r = 3 r_*$ (orange), which is exiting the domain at $\theta = 52^{\circ}$ for model B and $\theta = 64^{\circ}$ for model B3. By comparison, the $r = 1.5 r_*$ (red) streamline has only shifted from $\theta = 38^{\circ}$ to $\theta = 40^{\circ}$. Globally, the wind has changed very little between these two models, with the wind opening angle changing by only $\delta \omega \approx 5^{\circ}$. The change in separation between the two streamlines has changed from $\Delta \theta \approx 14^{\circ}$ to $26^{\circ}$, significantly more than the global wind opening angle. 

In Fig. \ref{fig:outflow_avg}, (top panel) we plot the mass flux $\rho v_r$ (solid lines) and $\theta-$integrated mass flux 
\begin{equation}
\dot{m} = 2 \pi \int_0^{\theta} \rho(r_o,\theta') v_r(r_o,\theta') \ \sin \theta' d \theta',
\end{equation}
(dashed lines) as a function of $\theta$ at the outer boundary $r_o$, time-averaged over $900 \ \rm{s} \leq t \leq 1000 \ \rm{s}$ for models B (black), B4 (red), B3 (green), B2 (blue) and B1.5 (purple). The structure of the outflows is largely the same, with larger truncation radii runs transporting away more mass and having slightly wider opening angles. In the bottom panel, we plot the velocity $v_r$ (solid lines) and density $\rho$ (dashed lines) as a function of $\theta$ for the same models. Decreasing the truncation radii results in a decrease in the maximum outflow velocity. The angle of the velocity peak also increases as the truncation radii decreases, consistent with the angle of the momentum density peak which also increases. This corresponds to our expectation of the flow becoming more radial as the truncation radius is decreased and the driving force becomes more radial. 

We note that the velocity profile of model B and B4 are very similar, despite the total disc luminosity differing by $\sim 50 \%$. The fast-stream is formed from gas launched in the innermost parts of the disc $r \leq 4 \ r_*$. The local radiation field near the inner disc is the same for these models since B4 has $r_c = 4 \ r_*$. This results in similar fast-stream structures and nearly identical velocity profiles. The difference in the outflows appears for $\theta > 65^{\circ}$, where model B has a denser, slower wind than B4. Truncating the disc results in a less dense but faster wind and a decrease in mass flux of $\sim 45 \%$. The difference in outflows at larger inclination angles can be understood by considering the mass loading in the outermost parts of the flow. Beyond the truncation radius $r > r_c$ the vertical radiation force is weak, resulting in a decreased density. Once high enough above the disc, the gas is accelerated by the radiation field interior to the truncation radius (which are the same in both models). To first order the mass flux is determined by the total luminosity, which are approximately identical since it is dominated by the innermost parts of the disc. The truncated model B4 will therefore have a faster outflow velocity, owing to its decreased density due to suppressed mass loading.

In Fig. \ref{fig:mdot_summary}, we plot mass flux $\dot{M}$ at the outer boundary as a function of time (solid lines) and the time average over $200 \ \rm{s} \leq t \leq 1000 \ \rm{s}$ (dashed line) for models B (black), B4 (red), B3 (green), B2 (blue) and B1.5 (purple). The mass flux is non-stationary with deviations ranging between 16\% (model B) to 29\% (model B4). We so no clear trend between time variability of solutions and truncation radius, which tended to be $\sim 20\%$, irrespective of the model.

\subsection{Luminosity Scaling}
\label{sec:luminosity}
\begin{figure}
                \centering
                \includegraphics[width=0.45\textwidth]{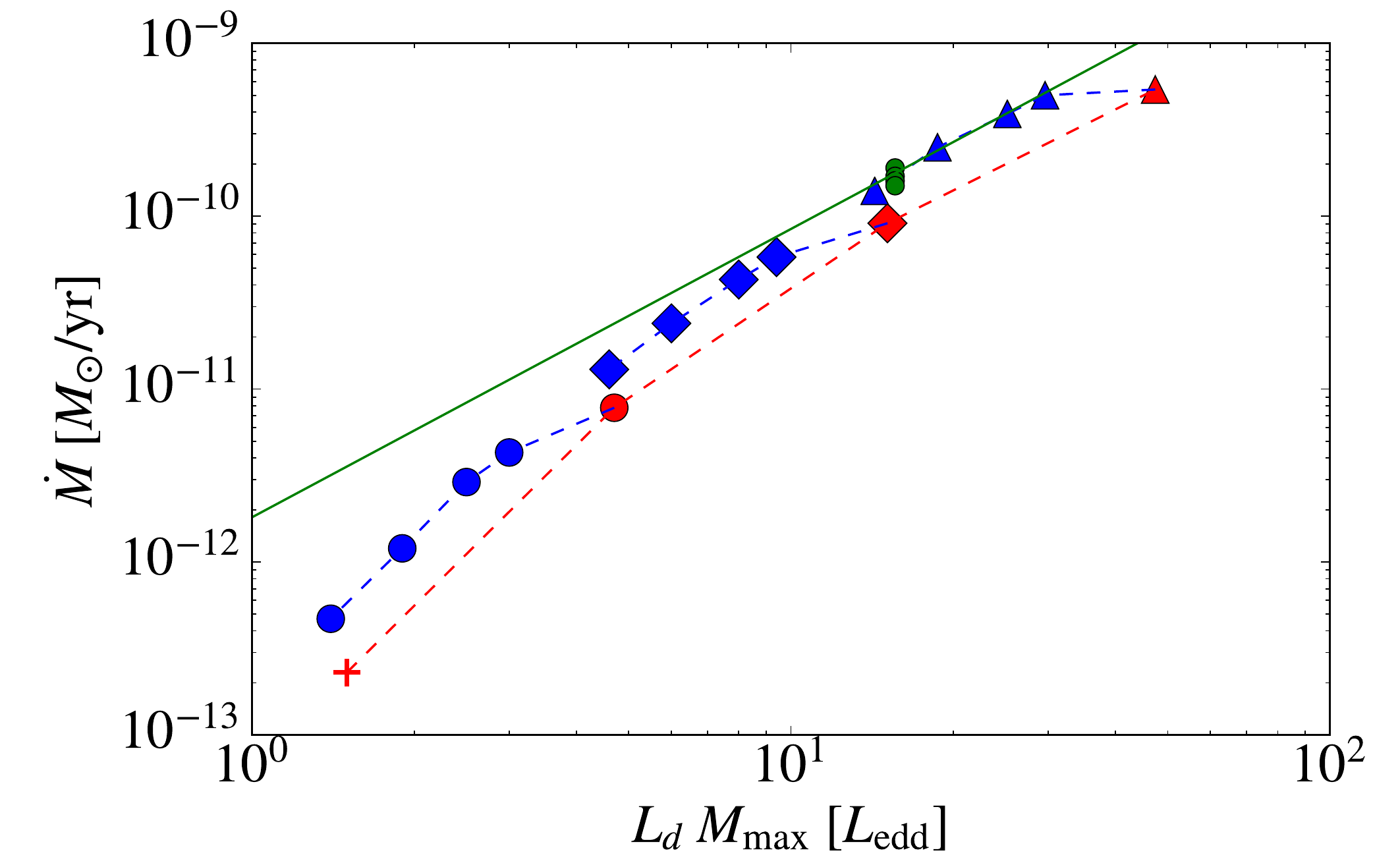}
                \includegraphics[width=0.45\textwidth]{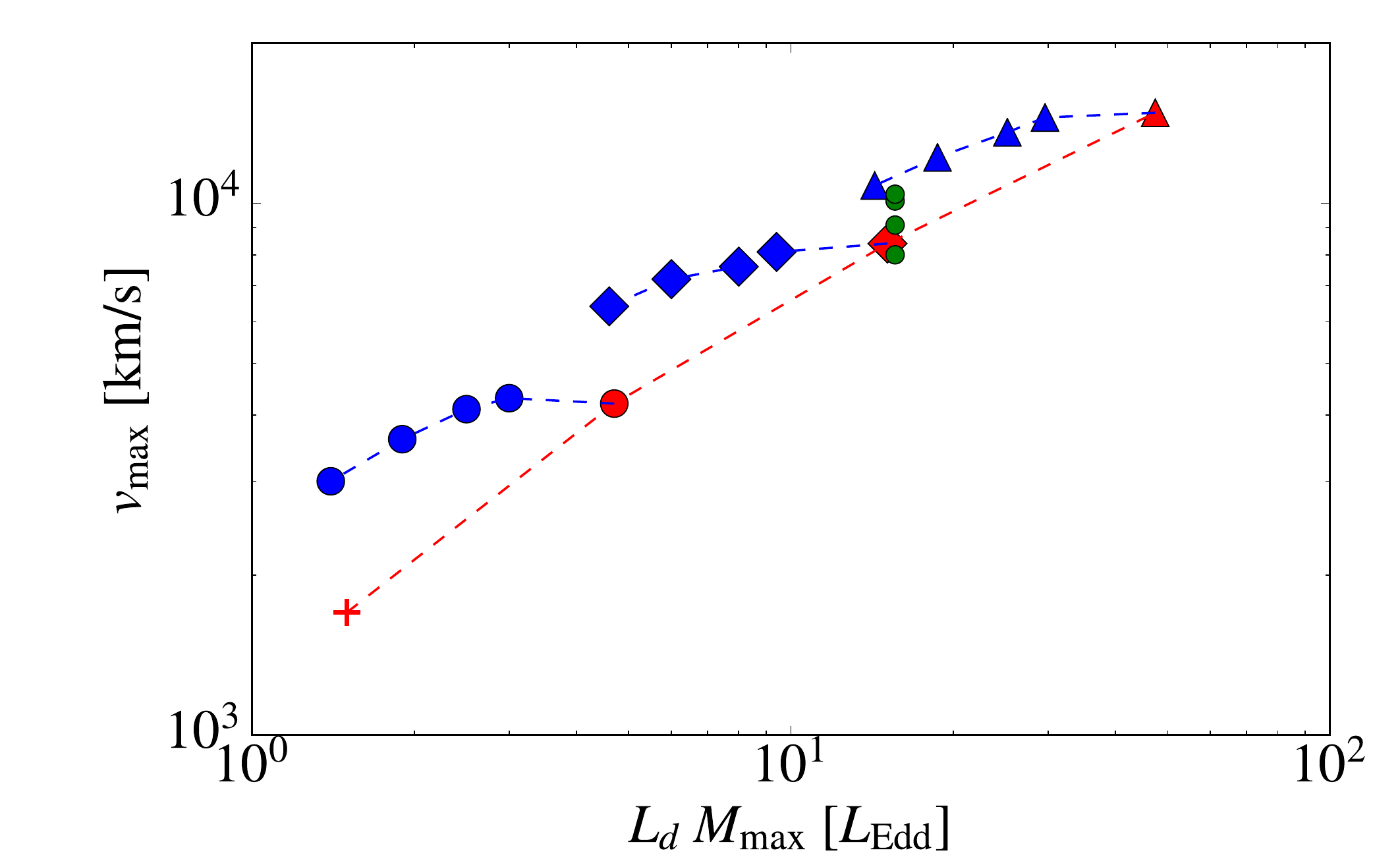}
        \caption{Global wind properties as a function of total system luminosity. Red symbols indicate Shakura-Sunyaev models, blue symbols the corresponding truncated disc models and green symbols the uniform disc models. The different symbol shapes indicate the Eddington parameter of the model, Model A (crosses), Model B (circle), Model P (diamond) and Model R (triangle). \textit{Top -} Mass flux $\dot{M}$ at the outer boundary as a function of total luminosity.  The Shakura-Sunyaev models display an increasing trend in the mass outflow, with a sharp cutoff when $ L_d \ M_{\rm{max}} \gtrsim 1$. At higher Eddington fractions truncated models approach the CAK solution (green line). \textit{Bottom -} Maximum velocity $v_{\rm{max}}$ at the outer boundary as a function of total disc luminosity. Maximum velocity tends to decrease as the truncation radius decreases but the dependence is weaker than with non-truncated models because the fast-stream originates from streamlines close to the inner disc edge.}
\label{fig:gloabl_vs_L}
\end{figure} 

\begin{figure}
                \centering
                \includegraphics[width=0.45\textwidth]{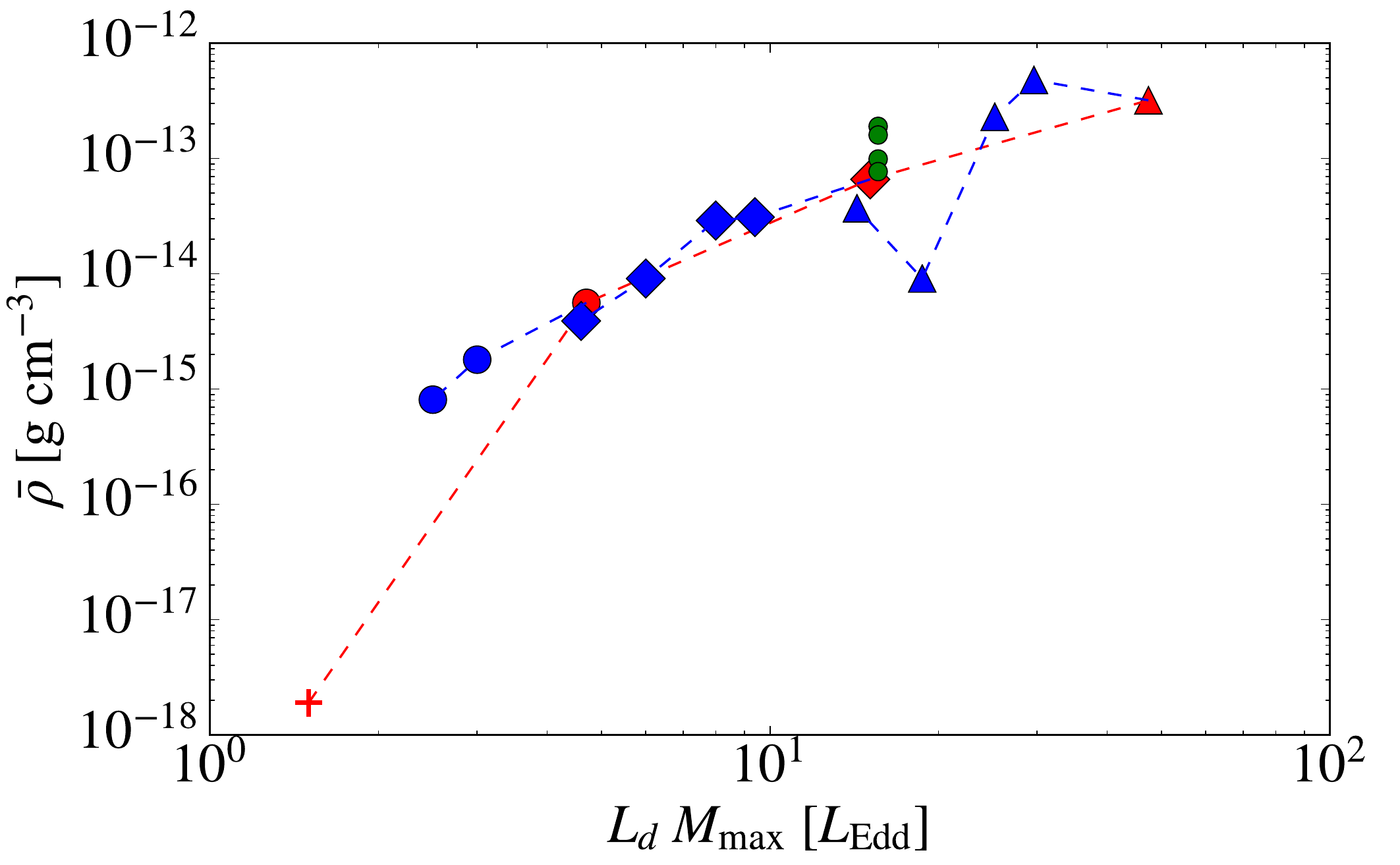}
                \includegraphics[width=0.45\textwidth]{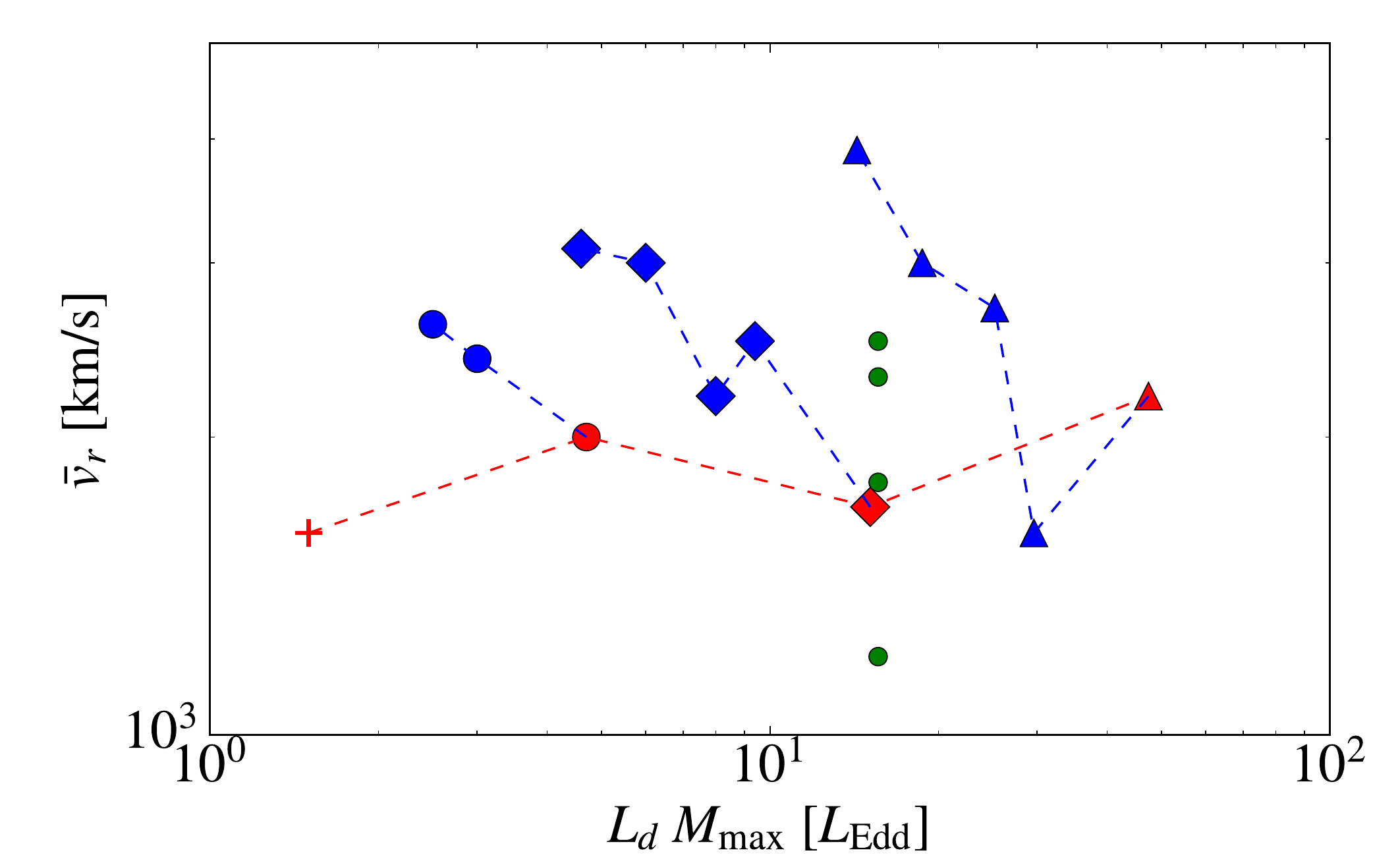}
        \caption{Fast-stream properties as a function of total system luminosity. Symbols have the same meaning as in Fig. \ref{fig:gloabl_vs_L} \textit{Top -} Average density $\bar{\rho}$ at the outer boundary in the fast-stream as a function of total luminosity. Truncated and non-truncated models show the same type of scaling with luminosity. \textit{Bottom -} Average outflow velocity $v_{r}$ in the fast-stream as a function of total disc luminosity. Non-truncated models are relatively luminosity indepenent, whereas truncated disc models show a slight increase in outflow velocity due to the fast-stream being slightly less dense.}
\label{fig:stream_vs_L}
\end{figure} 

\begin{figure*}
                \centering
                \includegraphics[width=\textwidth]{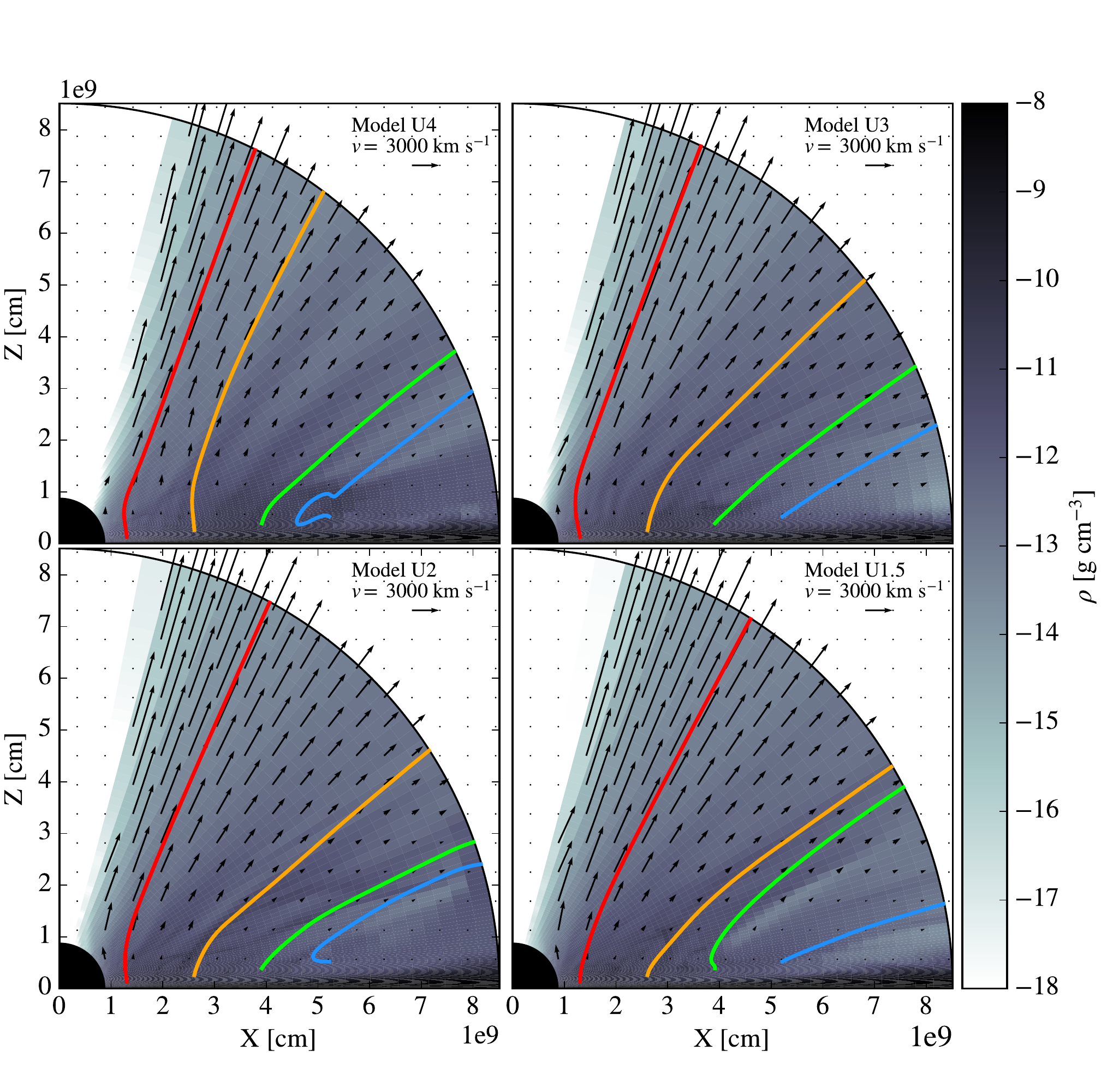}
        \caption{Time averaged density profile $\bar{\rho}$ (gray contours) and poloidal velocity field $\bar{\mathbf{v}}_p$ (black vectors) for $900 \ \rm{s} \leq t \leq 1000 \ \rm{s}$ for uniform intensity disc. This is the same as Fig. \ref{fig:time_averaged_winds_B} but for model U.}
\label{fig:time_averaged_winds_U}
\end{figure*} 

To first approximation, line driven wind properties are determined by the total system luminosity. Hence to better isolate geometric effects due to to the radiation field we compare outflow properties as a function of the total system luminosity.

In Fig. \ref{fig:gloabl_vs_L}, we plot global properties of the wind solutions, the time averaged mass flux $\dot{M}$ (top panel) and maximum outflow velocity $v_{\rm{max}}$ (bottom panel) as a function of total disc luminosity multiplied by the maximum force multiplier $L \ M_{\rm{max}}$, in units of the Eddington luminosity. Red symbols indicate Shakura-Sunyaev models, blue symbols the corresponding truncated disc models and green symbols the uniform intensity models. The different symbol shapes indicate the Eddington parameter of the model, Model A (crosses), Model B (circle), Model P (diamond) and Model R (triangle). When $L \ M_{\rm{max}} > 1$ the radiation force can overcome gravity to drive an outflow. Intuitively, this quantity can therefore be thought of as the imbalance between the radiation force and gravity.

We see that mass flux increases as a function of system luminosity for the non-truncated Shakura-Sunyaev models (dashed red line), with a particularly sharp increase near $L \ M_{\rm{max}} \gtrsim 1$. This is consistent with results from PSD98 (see their Fig. 8). For fixed Eddington parameter, we see that the mass flux decreases as the truncation radius decreases (blue dashed lines). The decrease in mass flux is less than predicted from the overall drop in luminosity however, with the non-truncated models (red dashed line) being a factor of $\sim 2$ below the truncated models with the same luminosity (dashed blue line). This is consistent with our understanding of the flow geometry that streamlines interior to the cutoff are largely unaffected by truncation and since these carry the bulk of the mass flux, the global outflow properties are therefore less affected by the truncation radius than one would predict from relying only on total disc luminosiy. We see that at higher luminosity and small truncation radius the mass flux approaches the scaling expected from CAK (solid green line). As the truncation radius is decreased, the radiation field becomes more spherically symmetric and we therefore expect better agreement with CAK. Our line driving treatment includes a maximum force multiplier which puts an upper limit on the radiation force as the gas becomes optically thin.  The CAK solution therefore \emph{overestimates} mass fluxes at lower luminosities when the force multiplier is saturated. 

The maximum outflow velocity of non-truncated models likewise shows a decrease as the total system luminosity decreases (red dashed line). As with the mass flux, decreasing the truncation radius leads to a decrease in maximum outflow velocity. However, we see this dependence is weaker, particularly when $r_c = 4 r_*$ and the outflow velocity is barely affected. The maximum velocity occurs in the fast-stream, so as with mass flux this part of the wind is not strongly affected by introducing a cutoff. When the truncation radius is small, $r_c = 1.5 r_*$, the velocity is $\sim 50\%$ greater than for non-truncated models with the same luminosity.

In Fig. \ref{fig:stream_vs_L}, we plot properties of the outflow bounded by streamlines with footpoints at $r = 1.5 \ r_*$ and $r = 3.0 \ r_*$. We show the average wind density $\bar{\rho}$ (top panel) and the density weighted outflow velocity $\bar{v}_r$ (lower panel) as a function of total disc luminosity $L \ M_{\rm{max}}$. Meanings of symbols are as in Fig. \ref{fig:gloabl_vs_L}. Density of non-truncated (red dashed line) and truncated (blue dashed lines) models scale approximately the same way for $L \ M_{\rm{max}} \gg 1$ with $\bar{\rho} \sim L^{4/3}$. Near $L \ M_{\rm{max}} \gtrsim 1$ the density drops sharply as we expect from the behaviour of the total mass flux.  

The velocity of non-truncated models are relatively luminosity indepenent (red dashed line), whereas truncated disc models show an increase in outflow velocity (dashed blue lines). Our analysis showed that mass flux scales approximately with the total luminosity. Therefore, since the density in the fast-stream is lower with decreased luminosity, the velocity must therefore increase. As with the maximum outflow velocity, this increase is approximately $50 \%$ above the non-truncated models. We conclude that truncating the disc allows us to launch \emph{faster} outflows for a given disc luminosity. 

Our analysis of winds driven by a truncated Shakura-Sunyaev radiation fields suggests that differences in outflow cannot be explained solely by differences in total disc luminosity. To further isolate geometric effects from those of system luminosity we study models with fixed luminosity and variable truncation radius. These models are denoted by the prefix U in Table \ref{tab:summary}. The Eddington fraction is chosen so that the total disc luminosity corresonds to the Shakura-Sunyaev luminosity $L_{SS}$ of a disc with Eddington fraction $1.18 \times 10^{-3}$ of the fiducial model (B) and $L M_{\rm{max}} = 15.6 \ L_{\rm{Edd.}}$.

In Fig. \ref{fig:time_averaged_winds_U}, we plot the time averaged density profile $\bar{\rho}$ (gray contours) and poloidal velocity field $\bar{\mathbf{v}}_p$ (black vectors) for $900 \ \rm{s} \leq t \leq 1000 \ \rm{s}$ for uniform luminosity disc models (The analogue to Fig. \ref{fig:time_averaged_winds_B} for model U). Uniform disc winds have a more vertical flow, since the radiation force at the base of the wind does not decrease as distance from the center increases. We can particularly see this with the innermost streamlines of U4, where they do not significantly diverge. For even the most truncated case, all streamlines exit the simulation domain, indicating that the flow is less variable at large radii than for truncated Shakura-Sunyaev models. The total mass flux still varies by $\sim 20\%$ however, suggesting that non-stationarity is a feature of disc winds, irrespective of the radial dependence of the radiation intensity. Unlike with the Shakura-Sunyaev disc, truncating the uniform disc has little effect on the global outflow properties. There is a slight decrease in the total mass flux ($\sim 25 \%$) and a slight increase in the maximum outflow velocity ($\sim 25\%$), but these are not evident from the momentum flux profiles. The shape of the outflow does not change noticably, in terms of the location of the fast-stream (shifted by only $3^{\circ}$) and by the peak in the velocity profile (unchanged). In the fast-stream we find that density decreases by a factor of $\sim 2$ as truncation radius decreases from $4 r_*$ to $1.5 r_*$. The velocity increases by a factor of $\sim 2$, as we would expect for the outflows to have constant mass flux. Our conclusion is the same as from the Shakura-Sunyaev cases, namely that truncating the disc allows us to have \emph{faster} outflows, for fixed disc luminosity. Truncated disc models have higher mass fluxes than non-truncated models with the same total luminosity.

\section{Discussion}
\label{sec:discussion}

We showed that though global outflow properties approximately scale with total system luminosity, radiation field geometry does play an effect. Truncating the disc can yield outflows with $\sim 2$ times greater mass flux and $\sim 50\%$ greater outflow velocity than a non-truncated disc with the same total disc luminosity. 

Geometry plays a crucial role in determining the ionization state of the wind. Altering the geometry of the UV radiation field changes the geometry of the flow, which alters the ionization state for a particular model of ionizing flux. Alternatively, with a fixed flow geometry, the ionization state changes in changing from a point to an extended ionization source. This points to the inevitable conclusion that correctly infering the radiation field geometry is critical in determining the strength and shape of the wind. The problem is further coupled by the fact that more accurate models of the radiation force due to lines is affected by the ionization state of the gas, which we neglect in the present work. These issues will have to be confronted if we are to properly study the question of self-shielding in AGN for example (see Matthews et al. 2016).

We showed that truncating the disc radiation alters the flow near the disc where radiation is suppressed. In the present work we have neglected both thermal and magnetic driving. Though these effects may be too weak to fully overcome gravity, our analysis suggests that this requirement may be too strong a condition as gas must only be lifted high enough above the disc so it may be radiated by the inner disc. It may be possible to radiatively drive outflows where $L M_{\rm{max}} \ll 1$ locally, provided that thermal/magnetic forces are strong enough to lift the gas high enough to overcome geometric foreshortening.   

Our work has shown that outflows are sensitive to the local radiation field, particularly near the disc where geometric foreshortening is important. An important next step is to compute the local UV intensity at every point in the wind, and use this as the source of line driving. This is different from our current setup, where frequency is integrated locally and this frequency integrated intensity is used to compute the intensity in the wind. If we maintain the assumption that the wind is optically thin to the continuum, intensity is time independent so this computation can be precalculated and simulations performed with no loss in computational efficiency.

Our model assumes the radiation field is sourced by a standard accretion disc. However, mass loss in the disc due to outflowing gas explicitly violates the assumption of mass and angular momentum conservation on which this model is based. It is important therefore to model the disc accretion so that the local radiation field can be computed self-consistently. The first step is to introduce an $\alpha-$viscosity as in Shakura \& Sunyaev (1973) and compute the local Eddington fraction as a function of the local accretion rate. This will not significantly increase computational time for the radiation force since we are already integrating over the full disc at every time step to compute the radiation force. The $\alpha$ prescription will add viscous terms to the momentum and energy equations, but this is less computationally expensive than the radiation force since it depends only on local velocity gradients and does not require integrating over the disc.

Beyond these models, such as fully resolving the magnetrorotational instability and computing disc accretion without relying on $\alpha-$viscosity, our current approach will no longer be consistent as the disc can no longer be considered optically thin and the key model assumption that the radiation field is emitted from the midplane breaks down. Such an approach is important because a time-varying accretion rate will lead to a time-varying radiation flux. Time-varying radiation fields have been shown to produce density and velocity perturbations at the base of line driven winds (Dyda \& Proga 2018c) and such an approach is necessary to understand disappearing BALs in AGN for example. This approach will require the use full radiation transfer, which we will leave to future work.

This work has focused on computing the correct UV flux, responsible for mediating momentum transfer between the radiation field and the gas. Equally important for line driving is correctly computing the internal state the gas. The ionization parameter in the wind depends on the radiation field of the X-rays. Determining the ionization state of the gas is required to self-consistently calculate the force due to line driving, as the force multiplier depends on $\xi$ (Dannen et al. 2018 in prep). Further we should go beyond the isothermal approximation and compute heating/cooling from the relevant SED (Dyda et al 2017), because temperature too can affect the number of available lines and thermal driving may alter the global outflow properties.       

\section*{Acknowledgements}
This work was supported by NASA under ATP grant NNX14AK44G. 
 

\label{lastpage}

\end{document}